\documentclass[fleqn,10pt]{wlpeerj}

\usepackage{listings}
\usepackage{longtable}
\usepackage{hyperref}

\title{Code4ML: a Large-scale Dataset of annotated Machine Learning Code}

\author[1]{Anastasia Drozdova}
\author[1]{Polina Guseva}
\author[1]{Ekaterina Trofimova}
\author[1]{Anna Scherbakova}
\author[1]{Andrey Ustyuzhanin}
\affil[1]{Department of Computer Science\\
  NRU Higher School of Economics\\
  Moscow, Pokrovsky Bulvar, 11 \\}

\corrauthor[1]{Ekaterina Trofimova}{etrofimova@hse.ru}

\begin{abstract}
Program code as a data source is gaining popularity in the data science community. Possible applications for models trained on such assets range from classification for data dimensionality reduction to automatic code generation. However, without annotation number of methods that could be applied is somewhat limited. To address the lack of annotated datasets, we present the Code4ML corpus.
It contains code snippets, task summaries, competitions and dataset descriptions publicly available from Kaggle \textendash~the leading platform for hosting data science competitions. The corpus consists of $\approx~2.5$~million snippets of ML code collected from~$\approx~100$ thousand Jupyter notebooks. 
A representative fraction of the snippets is annotated by human assessors through a user-friendly interface specially designed for that purpose. Code4ML dataset can potentially help address a number of software engineering or data science challenges through a data-driven approach. For example, it can be helpful for semantic code classification, code auto-completion, and code generation for an ML task specified in natural language.
\end{abstract}

\begin{document}

\flushbottom
\maketitle
\thispagestyle{empty}

\section*{Introduction}

\par In recent years, more and more tools for software development started using machine learning (ML) (\citet{allamanis_survey, engineering_survey}). ML systems are capable of analyzing (\citet{review_maintainability, vulnerability_prediction}), manipulating (\citet{program_repair, code_completion}), and synthesizing (\citet{code_generation_transformer, large_lm_synthesis}) code. However, even the most successful deep-learning models of the last few years (\citet{proglang_translation, codex}) require training on vast amounts of data before they can obtain good results.

\par There is a multitude of code datasets (\citet{iyer2018mapping, puri2021project}). Still, most of them are not domain-specific, which poses a challenge during the development of tools for specialized areas of software engineering because of domain shift (\citet{domain_shift_coined}). Moreover, generic datasets can lack examples, making it hard for the model to pick up on domain-specific patterns.

\par ML is one of the most popular software development areas without a domain-specific code corpus. The lack of such a dataset hinders the development of data science tools. In this paper, we introduce a Large-scale Dataset of Machine Learning Code (Code4ML) dataset, a corpus of Python code snippets, competition summaries, and data descriptions from Kaggle.

\par Our key contributions are the following: 
\begin{itemize}
    \item We present a large dataset of about 2.5 million Python code snippets from public Kaggle notebooks. Those snippets are enriched with metadata. 
    \item We propose a novel technique for ML code annotation based on a Machine Learning Taxonomy Tree that reflects the main steps of the ML pipeline. In addition, we provide an annotation tool that can help continue further markup of the dataset. 

\end{itemize} 

\par The rest of this paper is organized as follows. Section 2 contains an overview of prior art datasets. Section 3 includes a description of our dataset collection/annotation process. 
Details of human and machine-oriented reading of the dataset are described in section 4. Section 5 describes potential applications and research directions that the community can perform with the presented dataset. Section 6 concludes the paper.

\section*{Related work}

\par Several publicly available datasets for source code have been proposed for various code intelligence tasks. Some datasets, like CodeNet (\citet{puri2021project}) and POLYCODER`s dataset (\citet{xu2022systematic}), contain snippets from different programming languages. Others consist of code from one specific language: PY150 (\citet{raychev2016py150}) for Python, CONCODE (\citet{iyer2018mapping}) for Java, Spider (\citet{yu2018spider}) for SQL, etc. The source code is collected from GitHub (CodeSearchNet \citet{husain2019codesearchnet}) and Stack Overflow (CoNaLa \citet{DBLP:journals/corr/abs-1805-08949}), and from other platforms as well, such as Kaggle (\citet{2021kgt}). In \citet{DBLP:journals/corr/abs-2102-04664} CodeXGLUE is proposed, a machine learning benchmark dataset that contains 14 datasets of different sizes and in multiple programming languages.

\par Table \ref{table:overview} gives an overview of several datasets for Python since our corpus is also for Python. As we aim to study ML code, we focus on ML-related datasets. In \citet{agashe2019juice}, the authors provide the set of manually created high-quality Jupyter notebooks representing the programming assignments for classes. The notebooks consist of alternating NL markdown and code cells. The code is assumed to match the provided markdown description. The motivation of the JuICe dataset lies in the generation of the code snippet by the natural description of the Jupyter notebook cell using the prior information from the notebook. Boa (\citet{biswas2019boa}) dataset represents a pool of data-science-related python files and the meta-information about the corresponding GitHub projects. In \cite{2021kgt} the authors present a KGTorrent dataset. It includes a full snapshot of publicly available artifacts of Kaggle, including Jupyter notebooks,  dataset descriptions, and forum discussions. 

\begin{table}[ht]
\centering
\makebox[\textwidth]{\begin{tabular}{p{4cm}p{2.5cm}p{2cm}p{2cm}p{2cm}}
\toprule
Dataset name  & Dataset size & Human-curated annotated data size & Data source & Natural description of the general task the code is written for \\ [0.5ex] 
\midrule
JuICe \citet{agashe2019juice} & $\approx~1.5M$ code snippets &  $\approx~4K$ code snippets & GitHub & - \\ 
\midrule
Boa \citet{biswas2019boa} & $\approx5M$ Python files & - & GitHub & - \\  
\midrule
KGTorrent \citet{2021kgt} &  $\approx~250K$ Jupyter notebook files & - &Kaggle & - \\ 
\midrule
\textbf{Code4ML}(ours) &  $\approx~2.5M$ code snippets &  $\approx~8K$ unique code snippets & Kaggle & \checkmark \\ 
\bottomrule
\end{tabular}}
\caption{\label{table:overview}Overview of some of the existing ML-related datasets for Python.}

\end{table}

\begin{figure}[!htb]
\minipage{0.48\textwidth}
  \includegraphics[width=\linewidth]{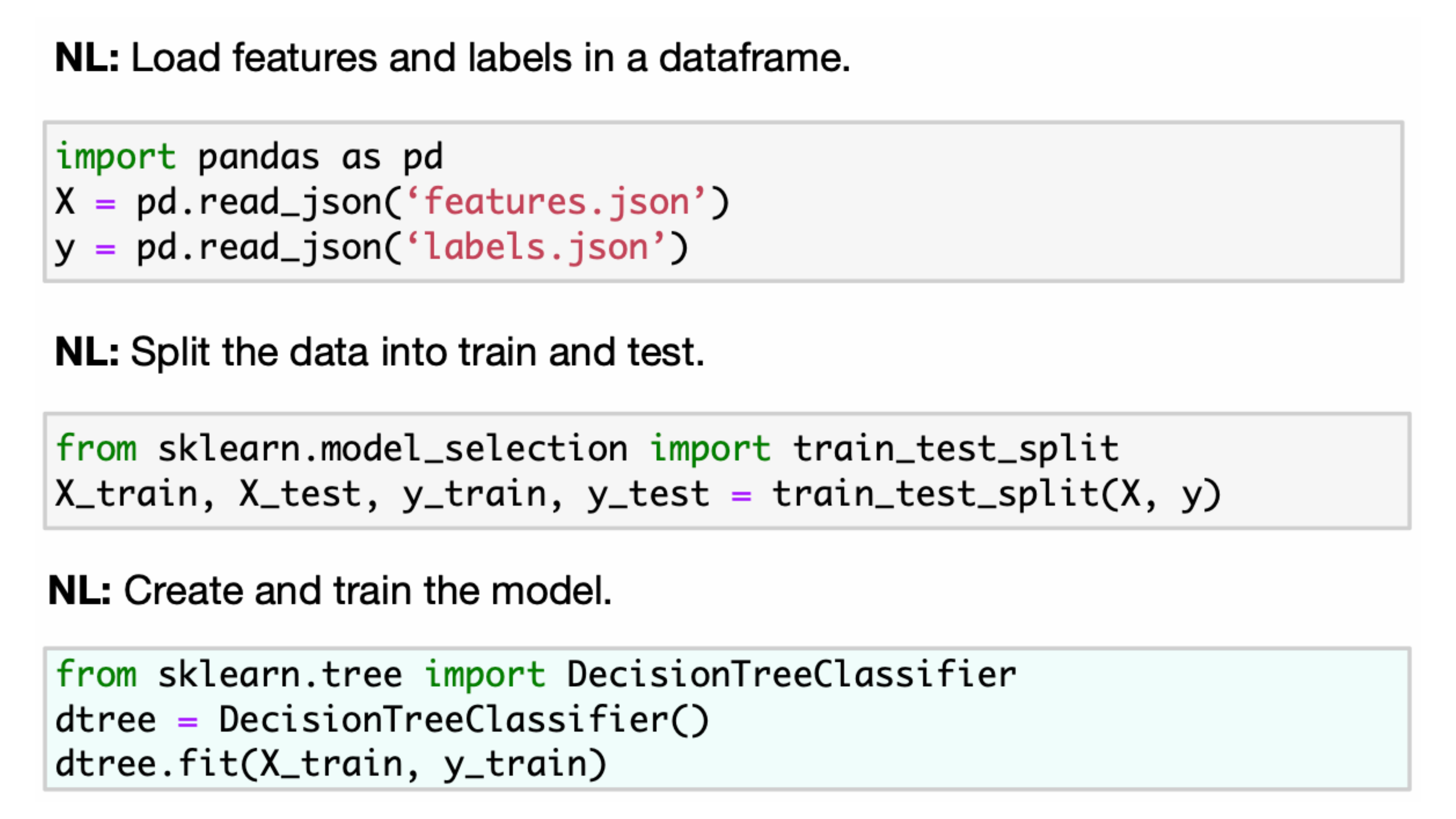}
  \caption{JuICE code snippets with the corresponding natural language description examples. Source: \citet{agashe2019juice}.}\label{fig:juice}
\endminipage\hfill
\minipage{0.48\textwidth}
  \includegraphics[width=\linewidth]{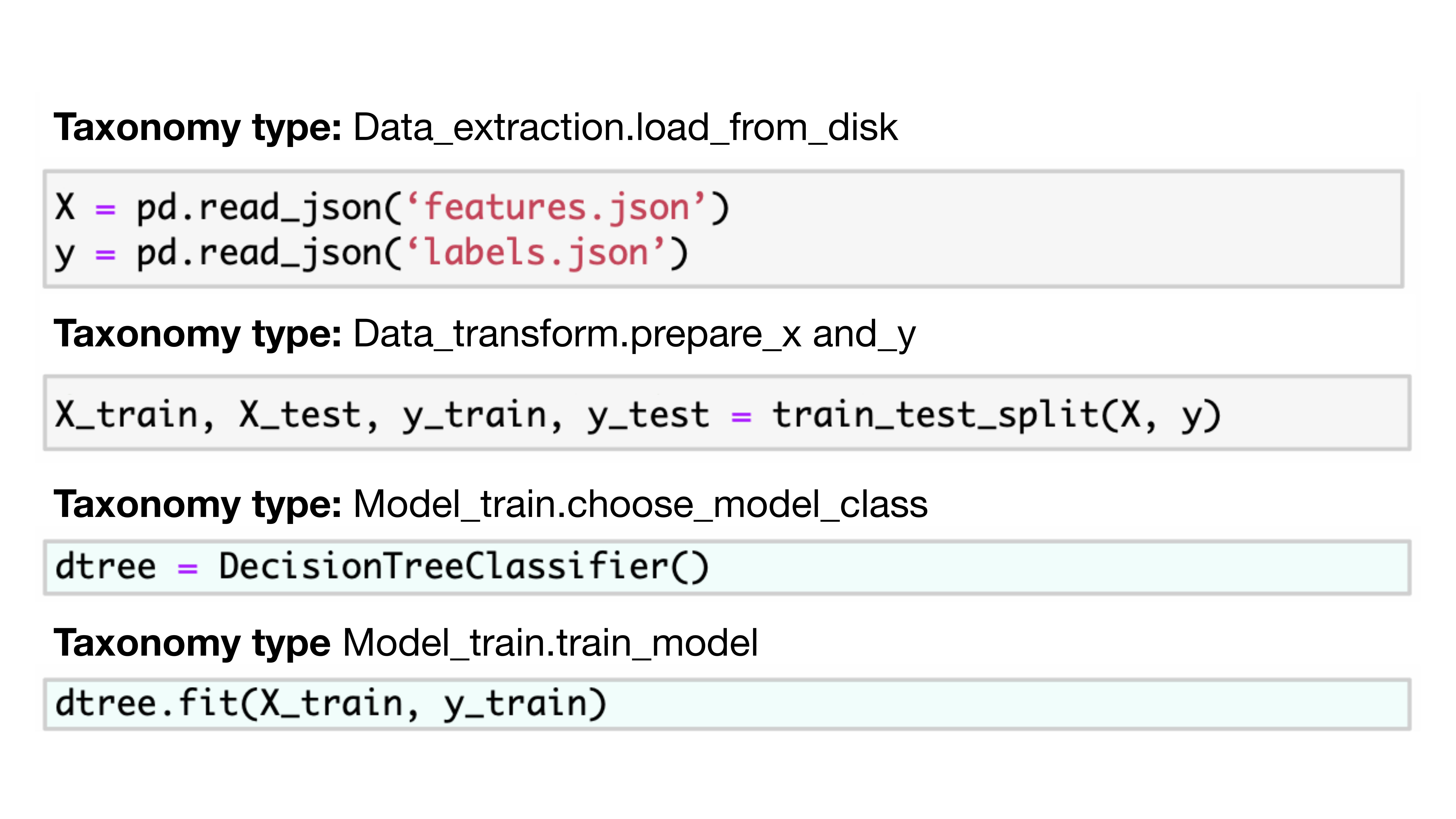}
  \caption{Code4ML code snippets with the corresponding taxonomy types examples. }\label{fig:Code4ML}
\endminipage
\end{figure}

Our work focuses on the Kaggle kernels (Jupyter Notebooks) as the sequential computational code cells designed to solve machine learning problems. We aim to reduce the dimension of the learning space by introducing a taxonomy tree once it is used as an annotation mark to notebook code cells. This annotation can be compared with the markdown describing the task of the code cell in JuICe dataset (see Figures~\ref{fig:juice} and \ref{fig:Code4ML}). However, unlike a markdown, our approach in the form of a taxonomy type is uniquely defined across all the snippets. We provide a set of $
\approx~8K$ human-curated annotated unique code snippets and a tool for the snippets' manual classification. Like KGTorrent, our corpus also contains information about Kaggle notebooks, corresponding datasets and competitions, and the competitions' natural descriptions. Thus, the whole ML pipeline, i.e., the sequence of the taxonomy tree vertices, is described by the human assessors.

\section*{Construction of Code4ML}

\subsection*{Collection and preprocessing of the dataset}
\label{sec:collect}

Kaggle is the most prominent platform for competitive data science. It curates the creation of data challenges and encourages users to publish their solutions in Jupyter Notebook kernels. A kernel is a sequence of code snippets and description blocks in a natural language. The code from the kernels is a good source of ML code since the users have to build their machine learning pipelines from scratch.

Kaggle provides an API for accessing the published kernels and competitions and an open dataset containing various metadata. Using the API, we collect the most popular kernels from the most popular competitions (i.e. with the highest number of teams).
We only consider kernels that use Python3 and have \href{https://www.apache.org/licenses/LICENSE-2.0}{Apache 2.0 license}.  

The collected kernels are further processed by the parser for code blocks and corresponding kernel id extraction.
Each code cell of the Jupyter notebook is considered a code snippet. We clean it up to ensure the collected code uniformity by removing broken Unicode characters and formatting the code to conform to the PEP8 standard. Also, personal information such as emails is not included in the snippets. 

Notebooks on Kaggle have many useful metrics. Users vote for notebooks with high-quality code. Another important notebook metric is the kernel result on the test set (Kaggle score). 

This metadata, as well as a number of kernel comments, are collected from Meta Kaggle. \footnote[1]{Kaggle's public data on competitions, users, submission scores, and kernels (\href{https://www.kaggle.com/datasets/kaggle/meta-kaggle}{meta-kaggle})}

\begin{figure*}[hbt]
	\centering
	\includegraphics[trim={0.2cm 0.7cm 1cm 0.7cm},clip,width=\textwidth]{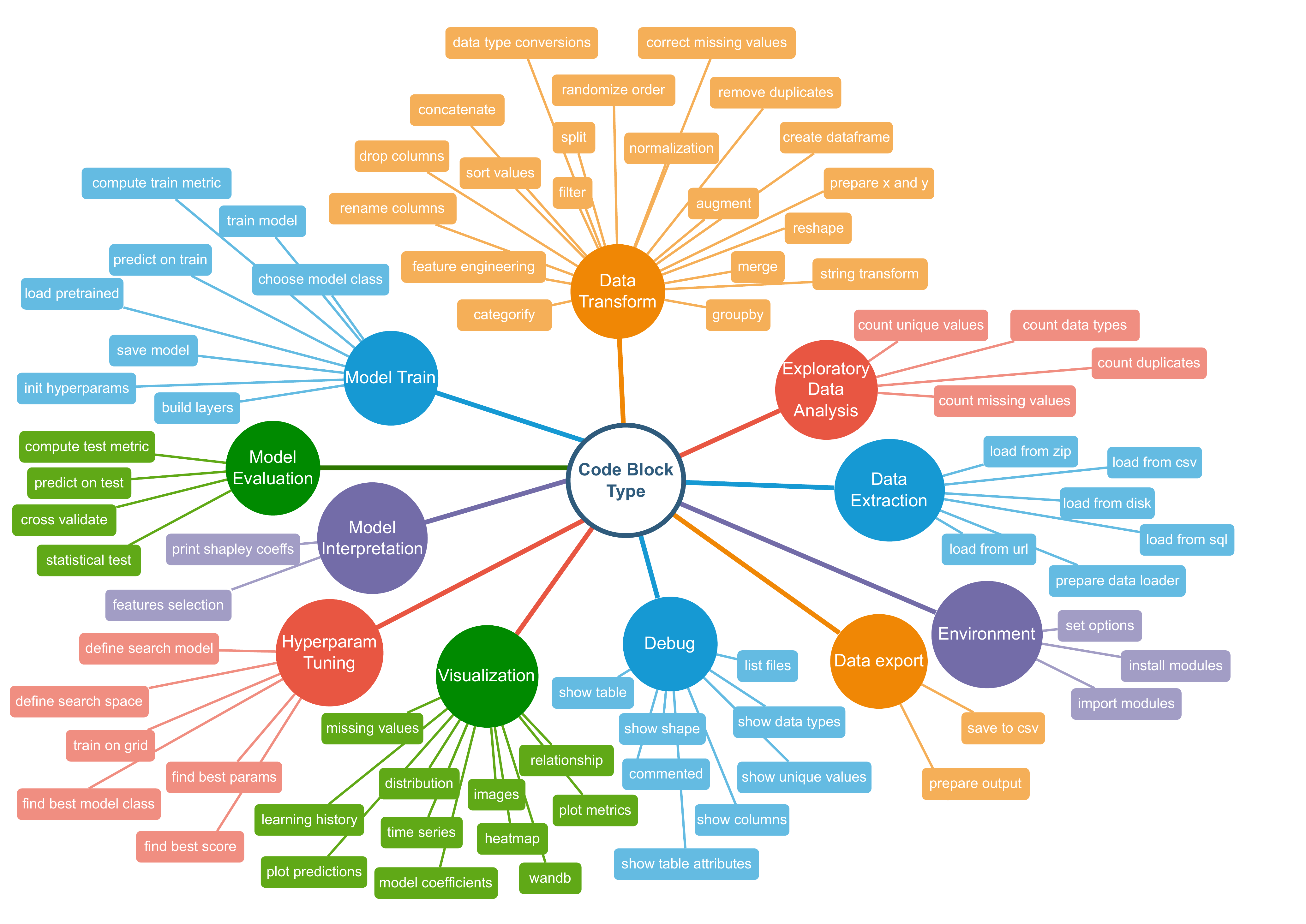}
	\caption{Machine Learning Taxonomy Tree.}
	\label{fig:knowledge_graph}
\end{figure*}

\subsection*{Taxonomy tree}
\label{sec:knowledge_graph}

Transformation of the Python code into conceptual pipelines describing the steps for performing ML experiments significantly reduces the amount of data required to train an ML model to analyse or generate the sequence. Almost any Jupyter Notebook can be translated into such a pipeline of repeating typical patterns. 

To describe code blocks from an notebook, we have developed a set of categories and combined it in a Taxonomy Tree. The tree has two levels: the upper level denotes a high-level category of an ML pipeline step. Each descendent vertex corresponds to a more specific action. The second-level vertices are called \textit{semantic types}. So, for example, semantic type \texttt{mising values} in \texttt{Visualisation} category represents an action of displaying missing values properties, such as quantities vs features. In contrast, \texttt{correct missing values} in \texttt{Data Transform} represents an action of filling it with a default value or removing the rows with missing values completely. There are 11 upper-level categories and $\approx~80$ lower-level classes. Figure~\ref{fig:knowledge_graph} illustrates the graph. Appendix~\ref{sec:code_snippets} shows examples of code snippets corresponding to different graph vertices.

Creating the ML Taxonomy Tree relies on data science standards such as CRISP-DM~\citet{shearer2000crisp} and ISOTR24029~\citet{ISOTR24029}, the experts’ experience in machine learning and data science.

\section*{Code4ML dataset structure}
\label{sec:dataset}

\begin{figure}[!htb]
\centering
  \includegraphics[trim={4.5cm 1cm 4.5cm 1cm},clip, width=0.9\linewidth]{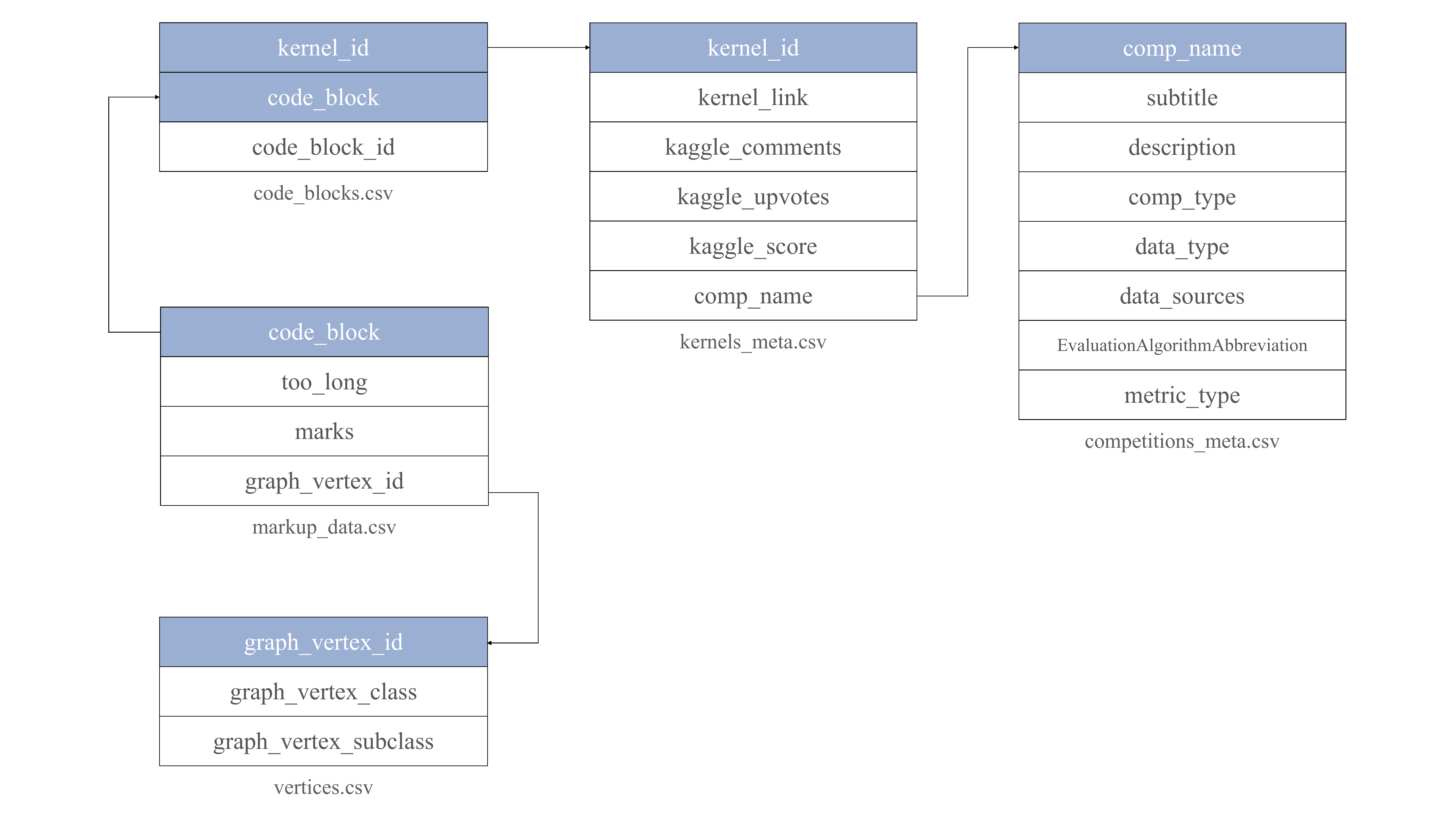}
  \caption{Code4ML corpus structure. Each table is stored in a separate file with a unique key. It is highlighted on the figure and used to reference its entries outside.}\label{fig:Code4ML_structure}
\end{figure}

The data is organized as a set of tables in CSV format. It includes several central entities: raw code blocks collected from Kaggle (\texttt{code\_blocks.csv}), kernels (\texttt{kernels\_meta.csv}) and competitions meta information (\texttt{competitions\_meta.csv}).  Annotated code blocks are presented in a separate table \texttt{markup\_data.csv}. Each code block is associated with a semantic type assigned to it by an external assessor. A Dictionary of semantic types is stored in the table \texttt{vertices.csv}.

Code snippets information (\texttt{code\_blocks.csv}) can be mapped with kernels metadata via \texttt{kernel\_id}. Kernels metadata is linked to Kaggle competitions information through \texttt{comp\_name} (figure~\ref{fig:Code4ML_structure}). To ensure the quality of the data \texttt{kernels\_meta.csv} includes only Jupyter Notebooks with an non-empty Kaggle score. The data is published online at Zenodo platform (\cite{anonymous_authors_2022_6607065}).

\begin{figure}[!htb]
\centering
  \includegraphics[trim={0cm 0cm 0cm 0cm},clip, width=0.9\linewidth]{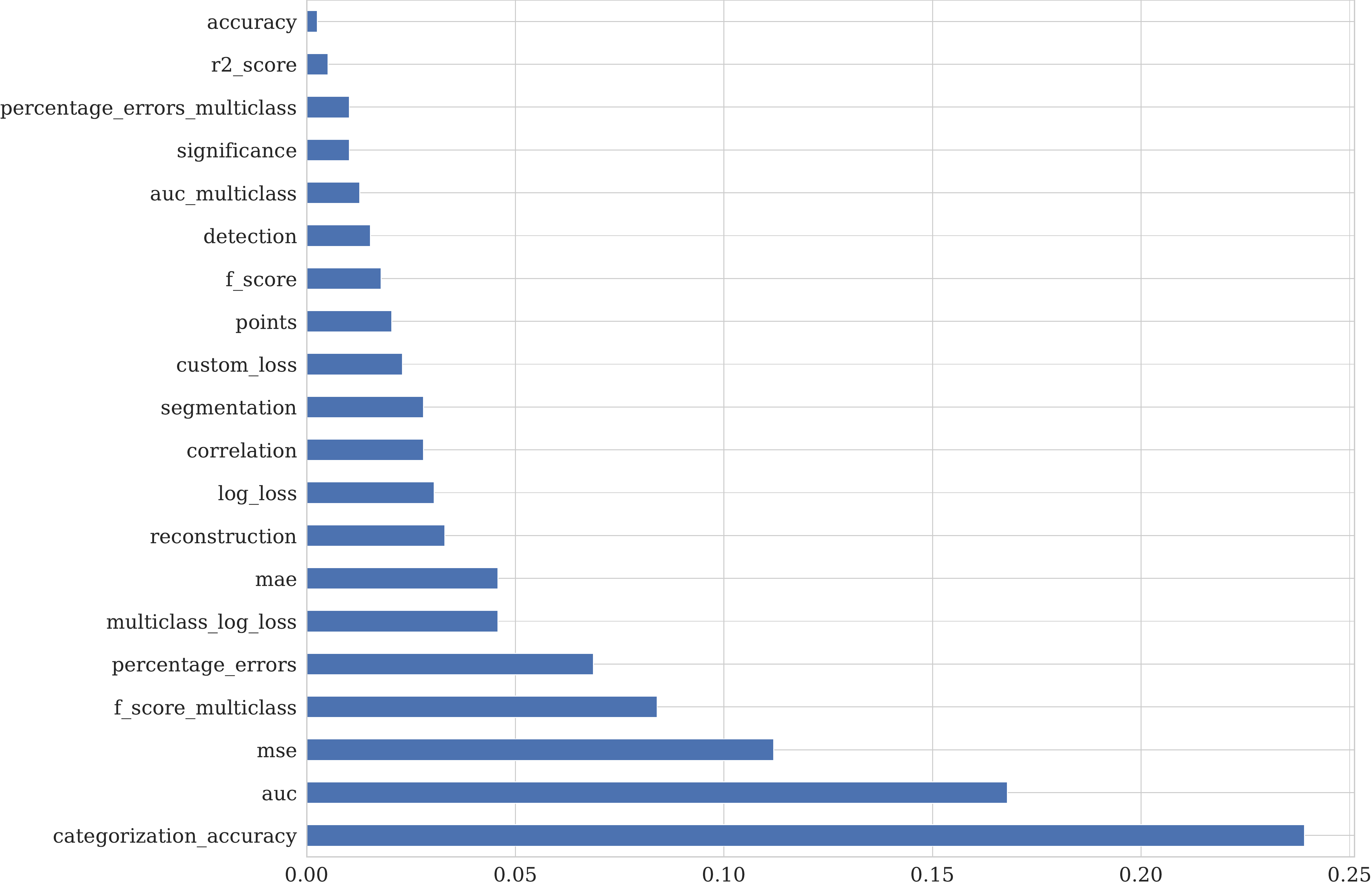}
  \caption{Distribution of the competition's metric type.}\label{fig:metric_type}
\end{figure}

Each \textit{competition} entry has the text description and metadata, reflecting competition, dataset characteristics, and evaluation metrics. 
\texttt{EvaluationAlgorithmAbbreviation} is collected from Meta Kaggle and provides additional informatoin on competitions and notebooks. \texttt{EvaluationAlgorithmAbbreviation} has 92 unique values, which make it difficult to filter the kernels by scores concerning the metric. To tackle it, we group \texttt{EvaluationAlgorithmAbbreviation} into 20 classes reflected in the \texttt{metric\_type} column. Figure~\ref{fig:metric_type} shows the distribution of the \texttt{metric\_type}. The description of each class is provided in appendix~\ref{sec:metric_type}.

The dataset for the corresponding competitions can be downloaded using Kaggle API: \texttt{kaggle competitions download -c data\_source}, where \texttt{data\_source} is the name of the dataset at Kaggle.

The \textit{code\_blocks} entry includes the code snippet, the corresponding kernel id and code block id, which is the index number of the snippet in the Jupyter Notebook.

The corpus contains 107\,524 notebooks. Almost a quarter of those (23\,104) are assigned to competitions. Thus, 625\,125 snippets belonging to those notebooks have a kernel score value.


We use a web form for manual sorting of code snippets into semantic classes described in Section \ref{sec:knowledge_graph}.\footnote[2]{Additional labelled data is always welcome. You can participate at \href{https://nl2ml-form.coresearch.club/}{https://nl2ml-form.coresearch.club/}. Please, take into attention that the registration of assessor needs to be approved by the Code4ML team members.}
The form allows marking code snippets according to their semantic type described in Section \ref{sec:knowledge_graph} as well as cleanliness and the kind of data (i.e., table, image, etc.) To specify the markup confidence level in the resulting class, one should choose the corresponding value of $\text{marks}$ (from 1 to 5). The $too\_long$ flag denotes the purity of the snippet to be marked up. The flag should be set if the cell code can not be unambiguously attributed to a single semantic type, i.e., it contains many different semantic types. One can find the detailed markup rules in the appendix~\ref{sec:MARKUP_RULES}. \texttt{markup\_data.csv} includes data labelled by the Code4ML project team. The interface of the web form is shown in the appendix \ref{sec:webinterface}. All assessors must follow the markup rules.  

The \textit{markup} table contains the following fields: the id of the parent notebook, code snippet text, the boolean \textit{too\_long} flag, the assessment confidence score in the range from 1 to 5 (best), and the id of the snippet class chosen by the assessor. 

In total, assessors marked around $10\,000$ snippets (some snippets are similar across notebooks, after that, there are $\approx8\,000$ unique snippets). $\approx68\%$ of marked snippets got the highest confidence score (i.e., 5), while $\approx18\%$ and $\approx11\%$ got the confidence score equal to 4 and 3, correspondingly. 

In order to annotate the rest of the corpus, we provide the general assessment of the automatic code snippets labelling.

We use the manually labelled code snippets for training the basic models. The class distribution of the snippets can be found in the appendix \ref{sec:class_distr}. We report two metrics: accuracy and F1-score.

{Since the code block is a sequence of symbols, an encoding is required. We used frequency-inverse document frequency (\cite{tfidf}) as a vectorizer.}

\begin{figure}[!htb]
\centering
  \includegraphics[trim={9cm 0cm 9cm 0cm},clip, width=0.7\linewidth]{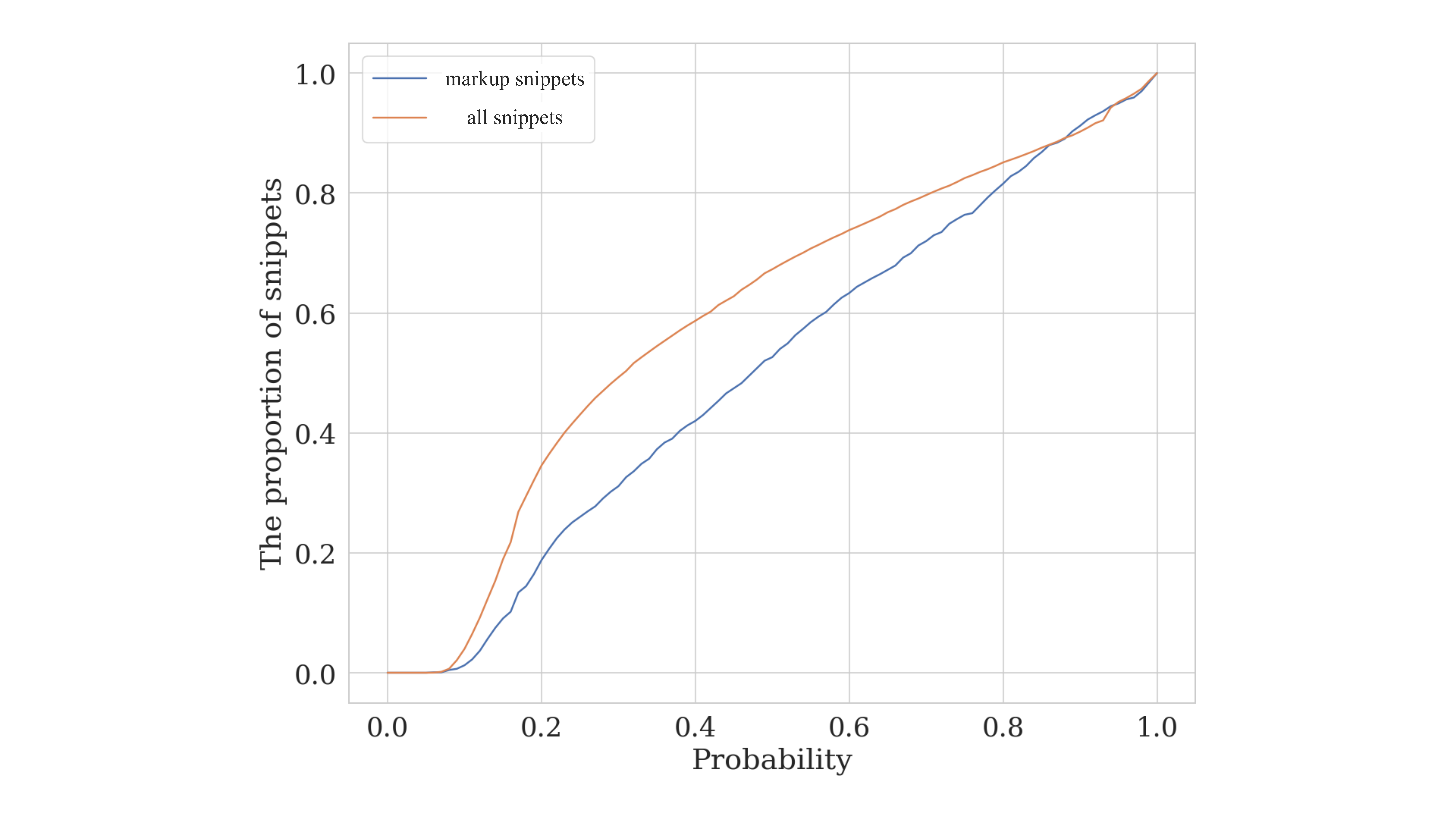}
    \caption{The valuation of the similarity of assessed and unassessed snippets. The plot lines show the cumulative distribution function (CDF) for the labelled (markup) and full (all snippets) samples depending on semantic class predicted probability.}\label{fig:snippet}
\end{figure}

We use a support vector machines (SVM) (\cite{bosertraining}) based models for snippets classification.
This method does not require much data for training, so this approach is used as a reference ML method. We apply SVM with different kernels: linear, polynomial and Radial Basis Function (RBF). The hyperparameters are selected based on cross-validation metrics on ten folds.  
The multiclass case is handled using a one-vs-all scheme (\cite{chang2011libsvm}). Details of the model training are available in the Appendix \ref{sec:hyperparameters}.

Figure~\ref{fig:snippet} illustrates the level of similarity between the manually assessed sample and the whole data. This plot shows the cumulative distribution function for the labelled and the full samples. The horizontal axis shows the prediction of a calibrated SVM classifier with a linear kernel, trained on 80\% of the labelled data. The probability ratio of the classes predicted by the model that does not exceed the specified threshold is then compared for the test part of the markup data (orange line) and the entire \texttt{code\_blocks.csv} table (blue line). Although the data in the whole dataset is not identical to the labelled data, one can see the closeness of the two lines, which allows us to conclude that the labelled sample is moderately representative.

The semi-supervised models (\cite{DBLP:journals/corr/abs-1911-04252}) for the snippets classification are applied to deal with the lack of manually labelled data.

First, 
linear kernel-based SVM model is trained on the marked-up dataset.
We collect the prediction of the trained model on the unlabelled part of the data. The predictions are further used as pseudo labels in combination with marked-up data to train a different SVM model with the RBF kernel.

\begin{table}[ht]
\centering
\begin{tabular}{l l l}
\toprule
& \multicolumn{2}{c}{Metrics} \\
\cmidrule(r){2-3}
Model  & F1-score &  Accuracy \\ [0.5ex] 
\midrule
SVM + Linear & 0.684 $\pm$ 0.024 & 0.691 $\pm$ 0.022 \\
SVM + Poly & 0.619 $\pm$ 0.021 & 0.625 $\pm$ 0.019 \\
SVM + RBF & 0.689 $\pm$ 0.022 & 0.625 $\pm$ 0.019 \\
\midrule
\midrule
SVM with 20 \% of pseudo labels & 0.831 $\pm$ 0.014 & 0.834 $\pm$ 0.014 \\
SVM with 40 \% of pseudo labels& 0.845 $\pm$ 0.016 & 0.851 $\pm$ 0.014 \\
SVM with 100 \% of pseudo labels& \textbf{0.872} $\pm$ 0.004 & \textbf{0.872} $\pm$ 0.004\\
\bottomrule
\end{tabular}
\caption{10-folds cross-validation performance of the baseline models for automatic data labeling.}
\label{table:all_results}
\end{table}

\section*{Downstream tasks}
The proposed corpus of the publicly available Kaggle code snippets, task summaries, competitions and dataset descriptions publicly enriched with annotation and useful metadata is a valuable asset for various data-driven scientific endeavors. 

As shown above, Code4ML can be used for a semantic code classification task, where each ML code snippet should be labelled as one of the taxonomy tree classes illustrated by the Figure \ref{fig:knowledge_graph}. This task helps summarize ML pipelines. For example, such a summary can serve as additional information or control input for the code generation model.  One can use the proposed baseline models as a starting point for the semantic ML code classification.

Code4ML also covers a lack of annotated data for ML code auto-completion and synthesis problems. Code completion is the most popular software development technique (\cite{autocompletion2006}) and hence is found in every major IDE. It can be used as a typing assistant tool for discovering relevant libraries and APIs.

Nevertheless, most existing code completion systems fail on uncommon completions despite their importance for real-world efficacy (\cite{autocompletion_fails}). Training a code completion model on domain-specific data can help determine the too rare patterns in generic code datasets and improve real-world accuracy.

Manually creating well-performing models requires significant time and computational resources, so automated machine learning (AutoML) serves to reduce the costs (\cite{automl_survey}). The task is usually tricky, and the state-of-the-art works in the field are focused on technical feature engineering and selection (\cite{automl_levels}). Creating problem-specific solutions from natural language specifications is still largely unexplored. For example, such specifications might contain domain-specific hints that the user would like to include in the pipeline/code. Still, the expression of such tricks in Python could be prohibitively expensive.

One of the problems in the AutoML field is the lack of relevant data for code synthesis. Training a generative model on a large-scale corpus of ML code like Code4ML can advance state-of-the-art automatic machine learning forward.

\section*{Conclusion}
\label{sec:conclusion}
This paper describes a novel Large-scale Dataset of annotated Machine Learning Code (Code4ML) containing ML code snippets in Python and corresponding ML tasks metadata.

The dataset contains problem descriptions from $\approx~400$ Kaggle competitions in natural language. It also includes more than 20 thousand public Python 3 notebooks representing machine learning pipelines for solving those competitions with the provided Kaggle score. Those notebooks comprise around $\approx~600$ thousand code cells. We propose a taxonomy graph to describe the code snippets as principal parts of the ML pipeline.  

The current version of the dataset does not cover the full scope of Kaggle ML code snippets and it can be easily extended in the future. 

Around ten thousand snippets have been manually labelled to the date. We developed a data markup web application that can help volunteers contribute to the extension of the markup dataset and eventually cover it entirely. Consequently, we warmly welcome any efforts from the community in this direction. 

We are confident that the Code4ML dataset can be helpful for various vital modern ML challenges, such as code classification, segmentation, generation, and auto-completion. Hopefully, it can also open up new venues for AutoML research.

\section*{Acknowledgement}
\label{sec:acknowledgement}

We want to acknowledge the considerable time and efforts spent annotating the Code4ML corpus by ([Names will be added upon review completion]).



\appendix
\onecolumn

\newpage
\section{Code snippets}
\label{sec:code_snippets}

\begin{longtable}{l l}
\toprule
\textbf{Semantic class} & \textbf{Example} \\
\midrule
Data\_Transform.drop\_column          &  \begin{lstlisting}[language=Python]
train_df.drop("Date", inplace=True, axis=1)
test_df.drop("Date", inplace=True, axis=1)
\end{lstlisting} \\
\midrule
Model\_Train.choose\_model\_class & \begin{lstlisting}[language=Python]
from sklearn import linear_model

reg_CC = linear_model.Lasso(alpha=0.1)
reg_Fat = linear_model.Lasso(alpha=0.1)
\end{lstlisting} \\
\midrule
Hyperparams.define\_search\_space & \begin{lstlisting}[language=Python]
parameters = {'lstm_nodes': [14,16,20],
              'nb_epoch': [50],
              'batch_size': [32],
              'optimizer': ['adam']}
\end{lstlisting} \\
\midrule
Visualization.distribution & \begin{lstlisting}[language=Python]
fig = plt.figure()
fig.suptitle("Algorithm Comparison")
ax = fig.add_subplot(111)
plt.boxplot(results)
ax.set_xticklabels(names)
plt.show()
\end{lstlisting} \\
\midrule
Data\_Transform.normalization & \begin{lstlisting}[language=Python]
scaler = MinMaxScaler()
df['revenue'] = scaler.fit_transform(
    df[['revenue']]
)
\end{lstlisting} \\
\midrule
Model\_Train.train\_model & \begin{lstlisting}[language=Python]
rfr = RandomForestRegressor(
    n_estimators=200, 
    max_depth=5, 
    max_features=0.5, 
    random_state=449,
    n_jobs=-1
)
rfr.fit(x_train, y_train)
\end{lstlisting} \\
\bottomrule
\caption{Examples of code snippets from Code4ML dataset} 
\end{longtable}

\newpage
\section{Metric type feature classes}
\label{sec:metric_type}

\begin{table*}[ht]
\centering
\begin{tabular}{p{4cm}p{7cm}p{2cm}}
\toprule
Class name & Description &  Aim  \\ [0.5ex] 
\midrule
accuracy & the share of the correct answers & maximisation \\ [1ex]
\midrule
r2\_score & coefficient of determination & maximisation\\ [1ex] 
\midrule
percentage\_errors\_multiclass & multiclass classification percentage error & minimisation \\ [1ex]
\midrule
significance & custom metrics reflecting  the predictions certainty & maximisation\\ [1ex] 
\midrule
auc\_multiclass & generalization of  ROCAUC  to a multiclass classification & maximisation\\ [1ex]
\midrule
detection & object detection metrics & maximisation\\ [1ex]
\midrule
f\_score & $F_{\beta}$-score metrics & maximisation\\ [1ex]
\midrule
points & reinforcement learning metrics & maximisation\\ [1ex]
\midrule
custom\_loss & custom loss metrics & minimisation\\ [1ex]
\midrule
segmentation & objects segmentation metrics & maximisation\\ [1ex]
\midrule
correlation & correlation metrics & maximisation\\ [1ex]
\midrule
log\_loss & logarithmic loss & minimisation\\ [1ex]
\midrule
reconstruction & reconstruction metrics & maximisation\\ [1ex]
\midrule
mae & mean average error metrics, e.g.  WMAE & minimisation\\ [1ex]
\midrule
multiclass\_log\_loss & logarithmic loss generalisation to multiclass classification  & minimisation\\ [1ex]
\midrule
persentage\_errors & percentage error metrics, e.g. RMSLE, mape  & minimisation\\ [1ex]
\midrule
f\_score\_multiclass & generalisation of f\_score to multiclass classification problems & maximisation\\ [1ex]
\midrule
mse & mean squared error metrics, e.g. mse, RMSE & minimisation\\ [1ex]
\midrule
auc & ROCAUC  & maximisation\\ [1ex]
\midrule
categorization\_accuracy & generalisation of accuracy to multiclass classification problems & maximisation\\ [1ex]
\bottomrule
\end{tabular}
\caption{Description of the \texttt{metric\_type} classes} 
\end{table*}

\newpage
\section{Markup rules}
\label{sec:MARKUP_RULES}

While marking up the data using the web form, one should take into account the following suggestions:
\begin{itemize}

  \item If code from only one semantic type is found in the snippet, $mark$ 5 is to be set;
  
  \item If the cell code can not be unambiguously interpreted, the $too\_long$ flag should be set up;

  \item If the $too\_long$ flag flag is set, the maximum possible $mark$ is equal to 4 (thus, the $too\_long$ flag and confidence 5 can not be set at the same time);
  
  \item If the snippet contains most of the code of one semantic type, but there is code of other types, then the type to which the most of the code belongs should be set with $mark$ equal to 4;
  
  \item If the snippet shares the same amount of code of different semantic types, the type that comes first with $mark$ equal to 3 should be set;
  
  \item If a sequence of operations for which semantic types are defined is applied to the data (e.g., sequence of $concatenate$ and $groupby$ methods) in one raw of code, then the semantic type of such a snippet will be the type of the last applied operation.
\end{itemize}

\newpage
\section{Interface of the web form}
\label{sec:webinterface}
\begin{figure}[ht]
    \centering
    \includegraphics[width=\textwidth]{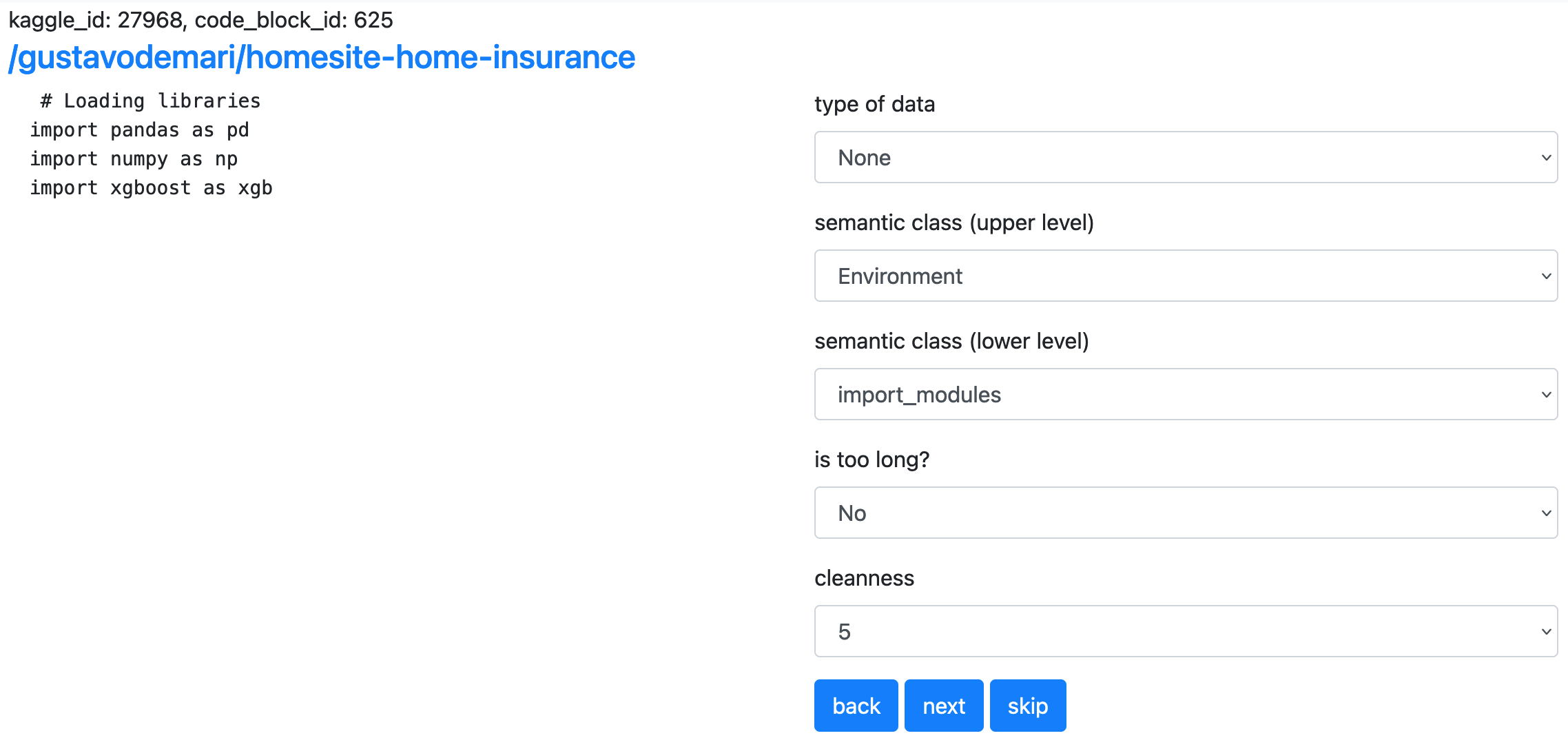}
    \caption{The interface of the web form. On the left there is an example of code snippet as well as the link to the original Kaggle kernel. On the right there are fields for manual labeling. Due to a large amount of options, the selection of semantic class is split into two parts.}
    \label{fig:webform_interface}
\end{figure}

\newpage
\section{Distribution of the manually labelled data}
\label{sec:class_distr}
\begin{figure}[ht]
    \centering
    \includegraphics[width=0.8\linewidth]{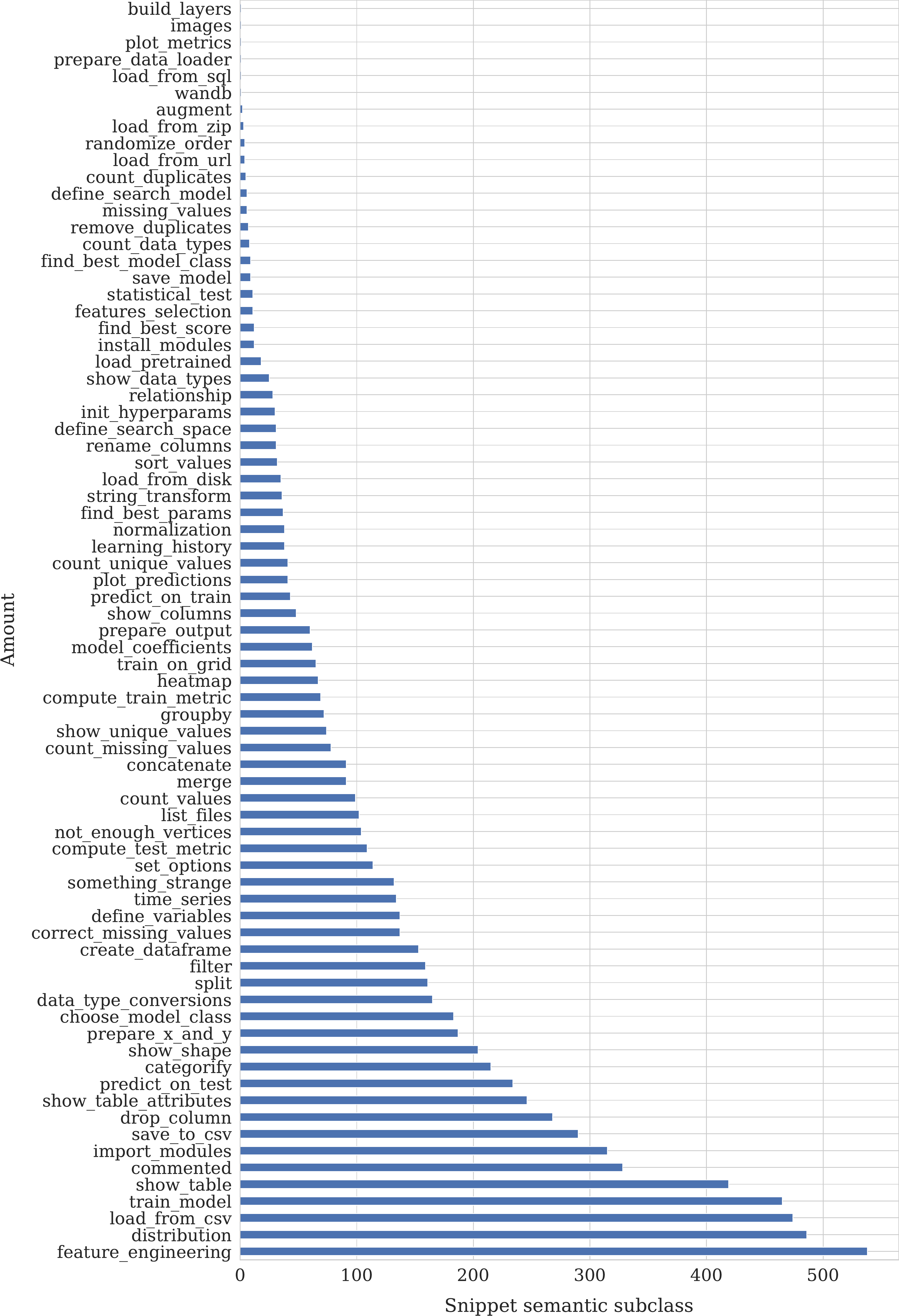}
    \caption{The dominated data type of the corresponding to the markup snippets competitions datasets is tabular. That leads to the imbalance in the semantic class distribution.}
    \label{fig:class_distr}
\end{figure}

\newpage
\section{Hyperparameters for automatic labelling}
\label{sec:hyperparameters}
\begin{longtable}{ l l r r }

\toprule
\textbf{Model} & \textbf{Hyperparameter} & \textbf{Type} & \textbf{Value} \\
\midrule
SVM + Linear          & C & numeric & 37.17 \\
                      & min\_df for TF-IDF & integer & 2  \\
                      & max\_df for TF-IDF & numeric & 0.31 \\
\midrule
SVM + Poly & C & numeric & 1.43 \\
           & Degree of poly kernel & integer & 3 \\
           & min\_df for TF-IDF & integer & 6  \\
           & max\_df for TF-IDF & numeric & 0.30 \\
\midrule
SVM + RBF & C & numeric &  8.71 \\
              & min\_df for TF-IDF & integer &  7  \\
              & max\_df for TF-IDF & numeric & 0.39 \\
\midrule
\midrule
Pseudo labels 20\% & C & numeric & 98.37 \\
                    & Kernel type & categorical & linear\\
                    & min\_df for TF-IDF & integer & 3  \\
                    & max\_df for TF-IDF & numeric & 0.53 \\
\midrule
Pseudo labels 40\% & C & numeric & 121.59 \\
                    & Kernel type & categorical & linear\\
                    & min\_df for TF-IDF & integer & 3  \\
                    & max\_df for TF-IDF & numeric & 0.41 \\
\midrule
Pseudo labels 100\% & C & numeric & 145.56 \\
                    & Kernel type & categorical & linear\\
                    & min\_df for TF-IDF & integer & 2  \\
                    & max\_df for TF-IDF & numeric & 0.26 \\
\bottomrule
\caption{Classification models hyperparameters. The hyperparameters for SVM models are selected by cross-validation on ten folds using Optuna \cite{optuna_2019}. 
The kernel can be Linear, Poly or RBF.
The regularization parameter C is selected from $[0.1, 1000]$.}
\end{longtable}

\bibliography{sample}

\end{document}